\documentclass[12pt]{article}
\usepackage[margin=1in]{geometry}
\usepackage{setspace}

\usepackage{float}

\usepackage{amsfonts}

\usepackage{amssymb}
\usepackage{latexsym}
\usepackage{graphicx}
\usepackage[english]{babel}
\usepackage[font={small}]{caption}

\usepackage{amsfonts}
\usepackage{latexsym}
\usepackage{graphicx}
\usepackage[english]{babel}
\usepackage{tikz}

\begin{document}
\input epsf

\def\p{\partial}
\def\h{{1\over 2}}
\def\be{\begin{equation}}
\def\bea{\begin{eqnarray}}
\def\ee{\end{equation}}
\def\eea{\end{eqnarray}}
\def\d{\partial}
\def\la{\lambda}
\def\eps{\epsilon}
\def\bb{\bigskip}
\def\mm{\medskip}
\newcommand{\dm}{\begin{displaymath}}
\newcommand{\edm}{\end{displaymath}}
\renewcommand{\b}{\tilde{B}}
\newcommand{\gm}{\Gamma}
\newcommand{\ac}[2]{\ensuremath{\{ #1, #2 \}}}
\renewcommand{\ell}{l}
\newcommand{\z}{\ell}
\newcommand{\newsection}[1]{\section{#1} \setcounter{equation}{0}}
\def\bb{$\bullet$}
\def\Qbar{{\bar Q}_1}
\def\QPbar{{\bar Q}_p}

\def\q{\quad}

\def\bn{B_\circ}

\let\a=\alpha \let\b=\beta \let\g=\gamma \let\d=\delta \let\e=\epsilon
\let\c=\chi \let\th=\theta  \let\k=\kappa
\let\l=\lambda \let\m=\mu \let\n=\nu \let\x=\xi \let\r=\rho
\let\s=\sigma \let\t=\tau
\let\vp=\varphi \let\vep=\varepsilon
\let\w=\omega      \let\G=\Gamma \let\D=\Delta \let\Th=\Theta
                     \let\P=\Pi \let\S=\Sigma

\def\h{{1\over 2}}
\def\t{\tilde}
\def\r{\rightarrow}
\def\nn{\nonumber\\}
\let\bm=\bibitem
\def\Kt{{\tilde K}}
\def\b{\bigskip}
\def\m{\medskip}

\let\p=\partial

\newcommand\blfootnote[1]{%
  \begingroup
  \renewcommand\thefootnote{}\footnote{#1}%
  \addtocounter{footnote}{-1}%
  \endgroup
}

\newcounter{daggerfootnote}
\newcommand*{\daggerfootnote}[1]{%
    \setcounter{daggerfootnote}{\value{footnote}}%
    \renewcommand*{\thefootnote}{\fnsymbol{footnote}}%
    \footnote[2]{#1}%
    \setcounter{footnote}{\value{daggerfootnote}}%
    \renewcommand*{\thefootnote}{\arabic{footnote}}%
    }

\begin{flushright}
\end{flushright}
\vspace{20mm}
\begin{center}
{\LARGE  Space cannot stretch too {\it fast}\daggerfootnote{Essay awarded fifth prize in  the Gravity Research Foundation 2025 Awards for Essays on Gravitation.}
 }

\vspace{18mm}
{\bf Samir D. Mathur$^{1}$ }

\blfootnote{$^{1}$ email: mathur.16@osu.edu }

\vspace{4mm}

\b

Department of Physics

 The Ohio State University
 
Columbus,
OH 43210, USA

\b

\vspace{4mm}
\end{center}
\vspace{10mm}
\thispagestyle{empty}
\begin{abstract}

We argue that  black holes microstates leave an imprint on the gravitational vacuum through their virtual fluctuations.  This imprint yields a power law fall off -- rather than an exponential fall off --  for the entanglement of planck scale fluctuations at different points. These entanglements generate an extra energy when space stretches too {\it fast},  since causality prevents a relaxation of these entanglements to their vacuum values.  We obtain semiclassical  dynamics for slow processes like star formation, but a radical departure from semiclassicality when a black hole horizon forms even though curvatures remain low everywhere.  This resolution of the information puzzle also implies an extra energy source at  the scale of the cosmological horizon, which may explain the mysteries of dark energy and the Hubble tension.

\end{abstract}
\vskip 1.0 true in

\newpage
\setcounter{page}{1}

\doublespace

Galilieo and Newton focused their  study on  material bodies, which were imagined to move through an inert, featureless vacuum. But further progress has indicated that  unlocking the secrets of nature requires a better understanding this vacuum itself. Einstein's gravity taught us that the vacuum has a {\it shape};  small ripples in this shape describe  gravitational waves. Quantum field theory taught us that the vacuum is full of {\it virtual particles},  which emerge as the real particles of Hawking radiation when space stretches during gravitational collapse.

\begin{figure}[h]
  \begin{minipage}[c]{0.5\textwidth}
    \includegraphics[scale=.8]{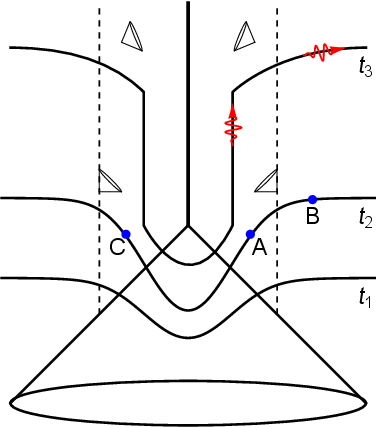}
  \end{minipage}\hfill
  \begin{minipage}[c]{0.5\textwidth}
    \caption{
       Semiclassical picture of gravitational collapse, in Eddington-Finkelstein coordinates. A shell collapses to a point, generating a singularity. Spacelike slices in this geometry stretch more and more as time evolves; this stretching leads to the creation of entangled pairs and the resulting information paradox.
    } \label{figbh}
  \end{minipage}
\end{figure}

But this phenomenon of Hawking radiation leads to the information paradox \cite{hawking}, depicted in fig.\ref{figbh}.  The `good slices' in the classical black hole geometry have low curvature everywhere. Time evolution leads to a stretching of these slices, which in turn leads   to the production of entangled particle pairs and the consequent monotonic rise of entanglement between the radiation and the remaining hole.  The small corrections theorem has provided a rigorous version of this conflict: if usual semiclassical physics holds at leading order around the horizon, then there is no way to prevent this monotonic rise of entanglement \cite{cern}.  Thus if information is to be recovered in the radiation, then there must necessarily be a second mode of failure of the semiclassical approximation -- one at low curvatures, distinct from the failure  when curvatures reach the planck scale.

\b

In this essay, we will see how a further fundamental change to our picture of the gravitational vacuum will resolve this long-standing puzzle. The role of nonperturbative vacuum fluctuations has been noted in \cite{vecro,elastic,secret}, but a crucial recent insight \cite{review} about the entanglement structure of these fluctuations leads to the complete picture described below. Remarkably this resolution implies a failure of Einstein's equations at the scale of the cosmological horizon as well,  something that may account for the mysteries of dark energy and the Hubble tension. 

\m

{\bf Modeling the vacuum}

\m

To describe quantum field theory we take  a lattice of points,  assign a field variable (a spin, a scalar $\phi(x)$ or something else) to each point, and choose an interaction between the variables at neighboring points. Gravity requires an obvious modification: the number of lattice points must increase when space is stretched. With this modification we obtain quantum fields on curved space, which imply phenomena like Hawking radiation.

But this cannot be a complete description of  quantum gravity, since the states of quantum fields cannot reproduce the $Exp[S_{bek}(M)]$ microstates of black holes ($S_{bek}=A/4G$ is the Bekenstein entropy \cite{bek}). What are we missing in our lattice model?

First consider  electrons and positrons. These exist as real (i.e., on shell) particles with energy $mc^2$. As a consequence,  the vacuum with zero energy contains {\it virtual} fluctuations corresponding to these particles. But these particles also form a bound state -- the positronium -- with a  radius $\sim 1\, A^o$. Does the  vacuum wavefunctional  have virtual fluctuations of positroniums?

A moment's thought indicates that including virtual positroniums   would be overcounting.  The way the positronium  manifests itself in the vacuum wavefunctional of QED is through correlation between the positions of virtual electrons and  positrons.   
The configurations where the virtual electron and positron are $\sim 1 \, A^0$ apart have a slightly higher amplitude since the energy of such configurations is slightly lower.  Similarly, other bound states of the standard model like the Benzene ring lead to subtle correlations among  fluctuations of the elementary fields in the vacuum wavefunctional. 

\begin{figure}
\begin{center}
\includegraphics[scale=.40]{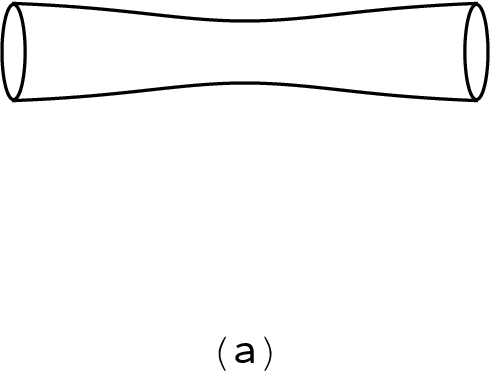}\hskip4em\includegraphics[scale=.40]{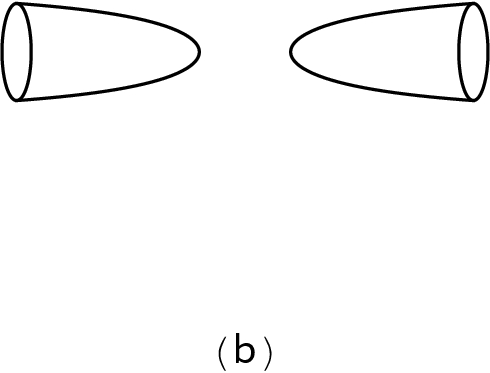}\hskip4em \includegraphics[scale=.40]{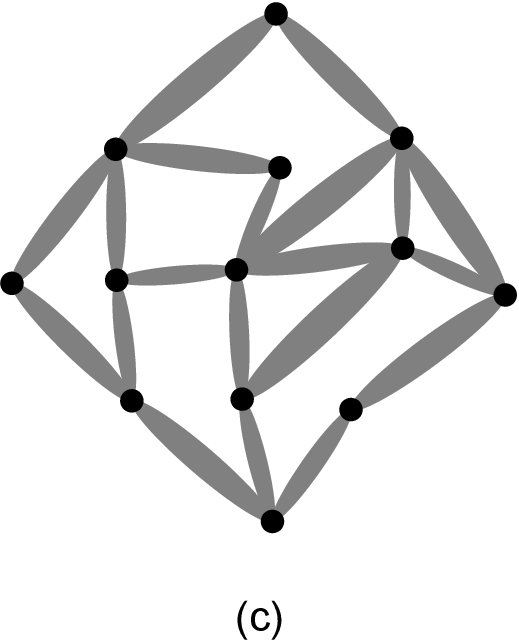}
\end{center}
\caption{(a) Small deformations of compact directions give scalar fields. (b) Larger deformations create planck scale solitonic objects like Kaluza-Klein monopoles. (c) Black hole microstates are `fuzzballls': complicated bound sets of such nonperturbative objects.}
\label{f3}
\end{figure}

We do not often worry about these subtle correlations because they are small.  But could there be another class of bound states that we have forgotten? Yes! There are $Exp[S_{bek}(M)]$ black hole microstates, for each value of $M$ with $0<M<\infty$. These microstates are bound states of the elementary objects in our quantum gravity theory, and must leave their imprint on the vacuum wavefunctional of quantum gravity through  correlations of the kind mentioned above.  

To understand this vacuum imprint, we must first know something about the on-shell microstates. In string theory, all the microstates which have been constructed have turned out to horizonless objects called {\it fuzzballs}: horizon-sized quantum balls made of the  fundamental objects in the theory -- strings and branes \cite{fuzzballs}.  A toy model of fuzzballs will be useful in extracting the essential properties we need.  In fig.\ref{f3}(a),(b)  we depict schematically the  compact and  noncompact directions. Small fluctuations of the compact directions (fig.\ref{f3}(a)) yield the light particles of the theory. But we can also have larger fluctuations that pinch off a compact direction to yield nonpertutbative solitonic objects like Kaluza-Klein monopoles (fig.\ref{f3}(b)).  A microstate for the hole of mass $M$ (depicted in fig.\ref{f3}(c)) has the form of a  bound state of these monopole-like objects, with overall radius $2GM+\epsilon$ ($\epsilon$ small).  The various fuzzball constructions bypass the usual no-hair arguments due to the special features of string theory, and  the fact that the compact directions are not trivially tensored with the noncompact ones in the interior of the fuzzball \cite{gibbonswarner,prevent}.    

 \m
  
{\bf Virtual black holes}

\m

While the on-shell fuzzball microstates of black holes are relatively well understood by now, our present goal is to model the effect of these bound states on the {\it vacuum} wavefunctional. Using our above schematic description of fuzzballs,  we can   develop a picture of what the quantum gravity vacuum should look like:

 \begin{figure}
\begin{center}
\includegraphics[scale=1.5]{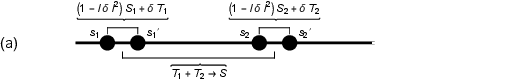}\\ 
\vskip2em
\includegraphics[scale=1.5]{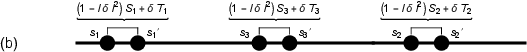}
\end{center}
\caption{ A 1-dimensional illustration: (a) Neighboring pairs of spins entangle to a state which is largely a singlet, with a small admixture of the triplet. The triplets from two such  pairs entangle to a singlet, giving a hierarchy of correlations between different length scales. (b) Expansion of space creates new pairs of spins in between, which must entangle with the previously existing spins in an appropriate way to yield the zero energy vacuum.}
\label{f2}
\end{figure}

(a) It is generally agreed that there will be violent fluctuations of spacetime at the planck scale. We can make this more concrete using fig.\ref{f3}. At planck scales we have not just fluctuations of massless fields as in fig.\ref{f3}(a), but also the topology changing fluctuations of fig.\ref{f3}(b) which generate the nonperturbative objects of string theory like KK-monopoles.  Fuzzballs involve these nonperturbative objects in a fundamental way, so to capture the effect of virtual fuzzballs we place a   monopole-like nonperturbative object at each site of our planck lattice. The nonperturbative objects of string theory form 
256-dimensional spin multiplets, but for simplicity we will attribute spin 1 to the object at each site.  In fig.\ref{f2}(a) we depict a 1-dimensional spacelike slice with  these nonperturbative objects at lattice sites depicted  by black dots. $s_1, s'_1$ are spin-1 objects at neighboring sites, and $s_2, s'_2$ are another such pair of objects at  sites further away.

(b) Recall that the on-shell positronium had a size $\sim 1\, A^o$, and its impact on the vacuum wavefunctional was to correlate fluctuations that are $\sim 1\,A^o$ apart. The on-shell fuzzballs are {\it extended} objects, with  radius $R\sim 2GM$ and $0<M<\infty$, so their imprint on the vacuum will be in the form of correlations across all length scales  $0<R<\infty$. In quantum theory, correlations between  two sites  are given by an {\it entanglement} between the states at these sites.  These correlations should fall off with distance, so we want a large entanglement between nearby sites and a smaller one between more distant sites.  

Such an  entanglement decreasing with distance  is schematically illustrated in the 1-dimensional picture of fig.\ref{f2}(a).  Let the spins at the neighboring sites $s_1, s'_1$ sites be entangled to produce  a state $
(1-|\delta|^2)S_1+\delta T_1$
 where $S_1$ is the singlet made out  of the two spins and $T_1$ is a triplet.  Similarly, the neighboring spins $s_2, s'_2$ are entangled in the state $ (1-|\delta|^2)S_2+\delta T_2$.  Finally, we entangle $T_1, T_2$ to make a singlet $S$.  With $\delta$ small, we find that each site is strongly entangled with a neighboring site, but there is also a smaller entanglement with sites further away.  
 
 In 3-dimensional space, we group the lattice sites into a set of hierarchical blocks of size $R_n\sim 2^n l_p$; a block at level $n+1$ is made by grouping $8$ neighboring blocks of size $R_n$. The spins in each block are entangled to make a spin-1 triplet, and the spins of $8$ neighboring blocks are again entangled to make a triplet of the larger block.  
 
 (c) The standard model has many bound states other than the positronium. A Benzene ring is an extended bound state made from the fundamental fields, and will leave its own imprint on the vacuum wavefunctional. But this imprint will be small, since the Benzene molecule is heavy.  Fuzzball states with $M\gg m_p$ are also heavy, so the effect of an individual fuzzball will be suppressed. But the {\it number} of these states is enormously large, and offsets the  suppression \cite{tunnel}. Thus we  expect that the entanglement in the vacuum  generated by the imprints of virtual fuzzballs will be very important.  

We quantify this entanglement by adopting the energy scales suggested by black holes. The vacuum with energy $E=0$ has the `optimal' entanglements between sites which yields the lowest energy state.  Suppose we have these optimal entanglements for all blocks upto those of radius $R_{max}$, but no entanglements at larger scales. The energy of such a  state would be higher than the energy of the vacuum. We set the extra energy in the region of size $R_{max}$ to be of the order suggested by the black hole relation $R\sim GM$: 
\be
\Delta E \sim {R_{max}\over G}
\label{en}
\ee
Note that the nonperturbative objects being entangled have mass $m\sim m_p$.  A quantum field with $m\sim m_p$ would have entanglements that decay exponentially over the distance scale $\sim l_p$. The postulate (\ref{en}) says that  the extended and numerous nature of black hole microstates leads to  an entanglement that falls off more slowly.  This fall-off may be characterized by the energy density $\Delta \rho\equiv \Delta E/R_{max}^3\sim 1/(GR_{max}^2)$, which falls as a power of the scale $R_{max}$ (rather than an exponential).

\b

{\bf Fast stretching}

\b

With the above model of the gravitational vacuum, we can explain our central idea: if spacetime  stretches too fast, then the semiclassical approximation breaks down.

Consider the spins in fig.\ref{f2}(a), entangled to yield the $E=0$ vacuum state. Now suppose space stretches, so that the spins $s_1, s'_1$ move away from the spins $s_2, s'_2$. New spins $s_3, s'_3$ must get created in between, as  in fig.\ref{f2}(b).  To achieve the vacuum on the stretched space, the entanglement  between the triplets $T_1, T_2$ must decrease, and an entanglement of strength $\delta$ must be created between $T_1$ and the new triplet $T_3$. 

The entanglement between  $T_1, T_2$ can change only if signals can be exchanged between these two groups of spins. If the stretching is slow, such signal  exchange is possible, and the  vacuum of fig.\ref{f2}(a) will evolve to the vacuum state in fig.\ref{f2}(b). But suppose the stretching is so fast that such signals {\it cannot} be exchanged. Then the entanglement of $T_1, T_2$ cannot break. By the monogamy of entanglement,  $T_1$ cannot then entangle with the newly created spin $T_3$ in the manner needed for the vacuum wavefunctional. We will end up with an extra energy (\ref{en}), with $R_{max}$ of order the separation between $T_1, T_3$. 

Consider  a  gas cloud that shrinks to form a star. Space stretches a little in this process, in the region inside the star. But light can traverse the star multiple times in the time taken for star formation, so the stretching is adiabatic and Einstein's equations will hold to an excellent approximation.  In a  gravitational  
wave the deformation of space cannot be considered `slow', but since the wave is transverse, we find that the volume of space does not change. Thus we do not create new spins of the type $s_3, s'_3$ in fig.\ref{f2}(b), and  will again find no violations of Einstein's equations \cite{review}.

But now consider the stretching of slices in gravitational collapse (fig.\ref{figbh}). Light signals cannot go from the point A inside the horizon to the point B just outside. In fact signals from point A cannot even reach point C which is at the same radius as A since light cones point purely `inwards' inside the horizon. Thus in a region of size $R_{max}\sim GM$ on the `good slice' $t_2$, the nonperturbative fluctuations of the vacuum will not be able to reach their vacuum entanglements.  These fluctuations will be in a state  with an extra energy $\Delta E\sim R_{max}/G \sim M$.  In the later slice $t_3$, space has stretched further inside the horizon, and the new stretched segment contributes a further energy $\Delta E \sim M$. Thus the `good slices' of the semiclassical geometry do not describe different foliations of a spacetime with mass $M$; instead, the energy along these slices rises by $\sim M$ when the time at infinity is advanced by $\Delta t\sim GM$. Since energy is fixed at the value $M$, the quantum gravity wavefunctional cannot evolve semiclassically along this foliation. 

\begin{figure}
  \begin{minipage}[c]{0.4\textwidth}
    \includegraphics[scale=.6]{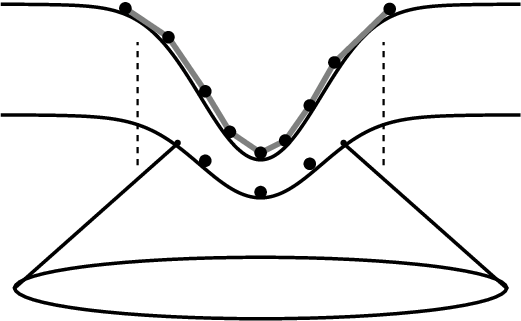}
  \end{minipage}\hfill
  \begin{minipage}[c]{0.6\textwidth}
    \caption{The fate of the collapsing shell, with our new picture of the gravitational vacuum. 
    Entanglements between planck scale fluctuations are unable to relax to their vacuum values for scales of order the  horizon radius. This leads to the creation of on-shell nonperturbative  objects (black dots) on the evolving slice, which then link up to create fuzzballs (upper slice) of the kind in fig.\ref{f3}(c).         
    } \label{figbh2}
  \end{minipage}
\end{figure}

If the evolution cannot proceed in the semiclassical manner of  fig.\ref{figbh}, what happens instead? The planck scale fluctuations have not been able to reach their vacuum entanglements, and thus show up as real excitations carrying the energy $\Delta E$ on the spacelike slice (fig.\ref{figbh2}). When these nonperturbative excitations link up in appropriate ways, we obtain one of the fuzzball states depicted in fig.\ref{f3}(c). There exist fuzzball states for each mass $M$,  so with the available energy the gravity wavefunctional is able to spread over this  space of mass $M$ fuzzball configurations. The resulting fuzzball has no horizon and radiates from its surface like a normal body, so there is no information paradox.

We can also see how the  fuzzball state that forms captures the information of the collapsing shell. Let the shell be composed of scalar quanta, described by the small fluctuations of a compact circle  with radius $R_c$, depicted in fig.\ref{f3}(a). The nonperturbative KK-monopole of fig.\ref{f3}(b) has a mass proportional to $R_c^2$. Thus we get a lower energy for KK-monopole fluctuations that are located close to a place where the scalar quantum reduces the size of this circle. Time evolution will favor such  configurations as the ones with lower energy, thus encoding the information of the collapsing shell into the wavefunctional over the space of allowed fuzzballs.

\m

{\bf Cosmology}

\m

Where else do we get `fast' stretching? In the expansion of the Universe, at the scale of the cosmological horizon $H^{-1}=\dot a/a$.  Einstein's equations will work well for distances $r\ll H^{-1}$,  since light can cross this scale $r$ multiple times over the expansion timescale $H^{-1}$.  This is no longer true at distances  $r\gtrsim H^{-1}$,   so entanglements across this scale need not be in their optimal vacuum form. By (\ref{en}), this implies an extra energy $\Delta E\sim {H^{-1}/G}$ over the volume of the cosmological horizon $(H^{-1})^3$; this  corresponds to an extra energy density $\Delta\rho\sim H^2/G\sim \rho_c$, where $\rho_c$ is the closure density. 

To see when we will induce this extra energy, we note a new twist that comes with cosmology.  At distances larger than the particle horizon $R_p(t)=a(t)\int_{t'=0}^t dt' /a(t')$, signals have not had time to travel to establish the entanglements of fig.\ref{f2}(a). So for distances $r>R_p$, fast stretching will not result in the entanglement problem of fig.\ref{f2}(b).  The constraint in the above para then says that $\delta \rho\ne 0$ only if $R_p\gtrsim H^{-1}$.  In \cite{elastic} it was noted that for the radiation phase $a\sim t^\h$ we have $R_p=H^{-1}$, while for the dust phase $a\sim t^{2\over 3}$ we have $R_p=2 H^{-1}$.  Thus in the radiation phase we will not get much extra energy from  fast stretching, which accords with the tight constraints on this phase from Big Bang Nucleosynthesis.  In the dust phase  on the other hand we {\it will} get an extra energy density $\Delta\rho\sim \rho_c$. This extra energy is of the correct order  to  account for the dark energy we see today.  The transition from radiation to dust induces a change in the entanglement profile of the nonperturbative virtual fluctuations; the energy associated with this  change might yield the Early Dark Energy that has been postulated at the radiation-dust transition to account for the Hubble tension.  

\m

 {\bf Summary and outlook}
 
 \m

Hawking's paradox demands that Einstein's gravity must fail  when a horizon forms, even though curvatures are low everywhere in a `good slicing' through the black hole geometry.  We have given an explicit  model for the gravitational vacuum which leads to such a failure, through the creation  of the energy (\ref{en}) when space stretches too {\it fast}.  

This energy induced by non-adiabatic expansion is reminiscent of the phenomenon of particle creation in curved space, which occurs when  vacuum  fluctuations of the type in fig.\ref{f3}(a) are unable to attain their optimal vacuum form due to quick stretching of space. But as we noted above, quantum fields describing these small fluctuations are unable to reproduce the microstates of black holes. In string theory these microstates are fuzzballs, composed of the nonperturbative objects of 
fig.\ref{f3}(b).   Black hole microstates are {\it extended} and {\it numerous}. Virtual fluctuations corresponding to these microstates leave an imprint on the vacuum in the form of a slowly falling entanglement between the virtual fluctuations of the nonperturbative  objects of fig.\ref{f3}(b). Since these objects are heavy, their fluctuations  relax quickly to their vacuum form when possible, screening such objects from view in slow processes like star formation. But in the black hole geometry the light cone structure disallows such relaxation, leading to a nucleation of fuzzballs in place of the traditional semiclassical stretching of slices. Fast stretching also occurs at the scale of the cosmological horizon, with potentially observable effects.

This picture of the gravitational vacuum also alters our thinking about the cosmological constant problem. String theory possesses a large number of compactifications with planck scale vacuum energy density, and  anthropic arguments are then invoked to select one with  small $\Lambda$. But in fuzzballs the compact directions are not trivially tensored with the noncompact ones. Thus virtual fluctuations of these black hole microstates should imbue the vacuum with a similar non-tensored structure. Instead of asking for the energy of a given compactification, we should ask for the energy of the state with optimal entanglements between the structures of the type in fig.\ref{f3}(b) created by the different ways of pinching off the compact directions.

The black hole horizon is a closed trapped surface, while  the cosmological horizon is a closed anti-trapped surface; the two are related by time-reversal.  In our picture, the same entanglements which resolve the information puzzle also give rise to dark energy. Thus we may have already seen quantum gravity  through astrophysical observations!


\newpage

 \section*{Acknowledgements}

This work is supported in part by DOE grant DE-SC0011726. I would like to thank Pierre Heidmann and Madhur Mehta for  helpful discussions.


\begin{thebibliography}{99}

  
 


 \bibitem{hawking}
  S.~W.~Hawking,
  Commun.\ Math.\ Phys.\  {\bf 43}, 199 (1975)
  [Erratum-ibid.\  {\bf 46}, 206 (1976)];
  S.~W.~Hawking,
  Phys.\ Rev.\  D {\bf 14}, 2460 (1976).
  
\bibitem{cern}
  S.~D.~Mathur,
  Class.\ Quant.\ Grav.\  {\bf 26}, 224001 (2009)
  [arXiv:0909.1038 [hep-th]].
 
  

\bibitem{vecro}
S.~D.~Mathur,
doi:10.1142/S0218271820300098
[arXiv:2001.11057 [hep-th]].

\bibitem{elastic}
S.~D.~Mathur,
Int. J. Mod. Phys. D \textbf{30}, no.14, 2141001 (2021)
doi:10.1142/S0218271821410017
[arXiv:2105.06963 [hep-th]].



\bibitem{secret}
S.~D.~Mathur,
Int. J. Mod. Phys. D \textbf{33}, no.15, 2440002 (2024)
doi:10.1142/S0218271824400029
[arXiv:2405.08945 [hep-th]].
  
\bibitem{review}
S.~D.~Mathur and M.~Mehta,
[arXiv:2412.09495 [hep-th]].

\bibitem{bek}
J.~D.~Bekenstein,
Phys.\ Rev.\ D {\bf 7}, 2333 (1973).
%


 
  
   \bibitem{fuzzballs}
O.~Lunin and S.~D.~Mathur,
  Nucl.\ Phys.\  B {\bf 623}, 342 (2002)
  [arXiv:hep-th/0109154];
O.~Lunin, J.~M.~Maldacena and L.~Maoz,
[arXiv:hep-th/0212210 [hep-th]];
I.~Kanitscheider, K.~Skenderis and M.~Taylor,
  arXiv:0704.0690 [hep-th];
 S.~D.~Mathur,
  Fortsch.\ Phys.\  {\bf 53}, 793 (2005)
  [arXiv:hep-th/0502050];\\
 I.~Bena and N.~P.~Warner,
  Lect.\ Notes Phys.\  {\bf 755}, 1 (2008)
  [arXiv:hep-th/0701216];
  B.~D.~Chowdhury and A.~Virmani,
  ``Modave Lectures on Fuzzballs and Emission from the D1-D5 System,''
  arXiv:1001.1444 [hep-th];
I.~Bena, S.~Giusto, R.~Russo, M.~Shigemori and N.~P.~Warner,
JHEP \textbf{05}, 110 (2015)
[arXiv:1503.01463 [hep-th]];
I.~Bena, S.~Giusto, E.~J.~Martinec, R.~Russo, M.~Shigemori, D.~Turton and N.~P.~Warner,
Phys. Rev. Lett. \textbf{117}, no.20, 201601 (2016)
[arXiv:1607.03908 [hep-th]].




  


\bibitem{gibbonswarner}
G.~W.~Gibbons and N.~P.~Warner,
Class. Quant. Grav. \textbf{31}, 025016 (2014)
doi:10.1088/0264-9381/31/2/025016
[arXiv:1305.0957 [hep-th]].

\bibitem{prevent}
S.~D.~Mathur,
Int. J. Mod. Phys. D \textbf{25}, no.12, 1644018 (2016)
doi:10.1142/S0218271816440181
[arXiv:1609.05222 [hep-th]].

\bibitem{tunnel}
S.~D.~Mathur,
Gen. Rel. Grav. \textbf{42}, 113-118 (2010)
doi:10.1007/s10714-009-0837-3
[arXiv:0805.3716 [hep-th]].
  S.~D.~Mathur,
  Int.\ J.\ Mod.\ Phys.\  D {\bf 18}, 2215 (2009)
  [arXiv:0905.4483 [hep-th]];
  P.~Kraus and S.~D.~Mathur,
  Int.\ J.\ Mod.\ Phys.\ D {\bf 24}, no. 12, 1543003 (2015)
  doi:10.1142/S0218271815430038
  [arXiv:1505.05078 [hep-th]];
  I.~Bena, D.~R.~Mayerson, A.~Puhm and B.~Vercnocke,
  JHEP {\bf 1607}, 031 (2016)
  doi:10.1007/JHEP07(2016)031
  [arXiv:1512.05376 [hep-th]].

 



\end{thebibliography}
\end{document}